# Beating Irrationality: Does Delegating to IT Alleviate the Sunk Cost Effect?

Philipp N. Herrmann

University of Paderborn

philipp.herrmann@wiwi.upb.de

Dennis O. Kundisch

University of Paderborn

dennis.kundisch@wiwi.upb.de

Mohammad S. Rahman[1]

University of Calgary

mohammad.rahman@haskayne.ucalgary.ca

**Abstract**

In this research, we investigate the impact of delegating decision making to information technology (IT) on an important human decision bias – the sunk cost effect. To address our research question, we use a unique and very rich dataset containing actual market transaction data for approximately 7,000 pay-per-bid auctions. Thus, unlike previous studies that are primarily laboratory experiments, we investigate the effects of using IT on the proneness to a decision bias in real market transactions. We identify and analyze irrational decision scenarios of auction participants. We find that participants with a higher monetary investment have an increased likelihood of violating the assumption of rationality, due to the sunk cost effect. Interestingly, after controlling for monetary investments, participants who delegate their decision making to IT and, consequently, have comparably lower behavioral investments (e.g., emotional attachment, effort, time) are less prone to the sunk cost effect. In particular, delegation to IT reduces the impact of overall investments on the sunk cost effect by approximately 50%.

---

[1] The authors are listed in alphabetical order.

# 1. Introduction

*"One of philosophy's oldest paradoxes is the apparent contradiction between the great triumphs and the dramatic failures of the human mind. The same organism that routinely solves inferential problems too subtle and complex for the mightiest computers often makes errors in the simplest of judgments about everyday events."* (Nisbett and Ross 1980)

*"We're wondering what a world looks like when there are a billion of these software agents transacting business on our behalf." – Dr. Steve R. White, IBM Research* (Chang 2002)

During the last decade, the role of information technology (IT) has evolved from being a decision aid to being a decision making artifact. Accordingly, nowadays, IT can not only support decision makers, but also make decisions on behalf of their owners (Chang 2002, Greenwald and Boyan 2001). Examples of these technologies include options for involving automated agents for bidding in online auctions (Adomavicius et al. 2009) or for trading in financial markets (Hendershott et al. 2011). Today, these agents are available at negligible marginal cost and can effectively act on behalf of their owners. As a result, for instance, in 2009 as much as 73% of all equity trading volume in the United States was executed by electronic agents (Mackenzie 2009). Not surprisingly, a significant literature has emerged, analyzing the design of these software agents, their performance in real market situations, and their effect on market outcomes (e.g., Guo et al. 2011, Hinz et al. 2011, Stone and Greenwald 2005). However, despite the widespread usage of these agents, the understanding of how delegating decision making to IT impacts different facets of decision making, especially decision biases, is still significantly lacking. Considering the economic importance of these decision biases (DellaVigna 2009), it is, nevertheless, critical to analyze the effects that the delegation of decision making to automated software agents has on the occurrence of decision biases.

Studies of decision biases have been featured in the literature for many decades (e.g., Camerer et al. 2004, Kahneman and Tversky 1979, Pope and Schweitzer 2011), including both laboratory and field research (overviews can be found in Camerer et al. 2004, DellaVigna 2009). One important challenge for researchers is to provide ways and means of how these biases can be alleviated or even avoided. Researchers from the information systems discipline have already made useful contributions in this area. Several laboratory experiments have shown that decision support systems (DSS) are an effective tool for alleviating some of these decision biases (e.g., Bhandari et al. 2008, Lim et al. 2000, Roy and Lerch 1996). However, none of these studies analyzed the role that automated software agents, which effectively replaces the decision maker for the delegated task and period, might have on the occurrence of decision biases in subsequent human decisions. In addition, there exists no evidence that the laboratory results associated with DSS and decision biases are transferable to real market situations. This handicaps academics as well as practitioners because many scholars are skeptical about the transferability of lab results to the field (Levitt and List 2008, List 2003). Consequently, we investigate whether or not IT can indeed alleviate a decision bias in real market transactions.

One frequently occurring decision bias is called the 'sunk cost effect'. It has been defined as *"a greater tendency to continue an endeavor once an investment in money, effort, or time has been made"* (Arkes and Blumer 1985). The sunk cost effect typically occurs in decision situations involving a chain of decisions (e.g., software projects, investments, exploration ventures, auctions) (Kanodia et al. 1989). In many of these situations, it is now possible to delegate parts of the decision making to IT (Chang 2002, Greenwald and Boyan 2001). Therefore, both researchers and practitioners would benefit from a better understanding of the impact of delegation to IT on the sunk cost effect. In this research, we have been fortunate enough to be able to address this issue by analyzing data obtained from a real market setting. In particular, we focus on the following research question: *Does the delegation of parts of the actual decision making to IT affect the proneness to the sunk cost effect in a real market situation?*

We consider explicitly if the delegation of decision making to IT decreases behavioral investments and if there is a sunk cost effect for these kinds of investments. While there is anecdotal evidence that the delegation of decision making to IT can reduce behavioral investments (Bapna 2003), we do not know of any paper which empirically investigates this issue. In addition, there is no clear picture of the impact of behavioral investments on the sunk cost effect. Some experimental studies have found a positive effect of behavioral investments on the sunk cost effect (e.g., Cunha and Caldieraro 2009, 2010, Navarro and Fantino 2009) while other scholars fail to replicate these experimental results (Otto 2010) or argue that there is no sunk cost effect for a behavioral investment such as time (Soman 2001).

The recent rise in online pay-per-bid auctions (e.g., beezid.com, bidcactus.com) gives us a real market setting to answer our research question.[2] The growth in popularity of this type of auction has attracted a lot of media attention (e.g., Harford 2010, Thaler 2009). For example, Thaler (2009) writes in the New York Times about a big pay-per-bid auction website: *"Here, the theory* [of the sunk cost effect] *is employed in some diabolically inventive ways."* In particular, we use a unique and very rich dataset provided by a German website offering pay-per-bid auctions with direct buy option.[3] This dataset includes detailed customer level bidding and transaction data from approximately 7,000 auctions between

---

[2] In September 2011, 5.5 million unique visitors visited pay-per-bid auction websites. This corresponds to 7.3% of the unique visitors on the biggest auction website worldwide – ebay.com (Platt et al. 2012).

[3] The website has requested to remain anonymous.

August 2009 and May 2010. Furthermore, our dataset allows us to distinguish on a bid level if a bid was placed manually or via an automated bidding agent. This unique feature of our dataset allows us to analyze the effect of delegating to IT on the proneness to the sunk cost effect.

We analyze normatively irrational decision scenarios of auction participants. In our empirical analysis, we find that the likelihood of making an irrational decision is significantly influenced by a participant's investments during an auction. Consistent with the sunk cost literature, this likelihood is an increasing function of the bidding fees paid in a specific auction: For example, a participant who invested aggregated bidding fees of $20 is much more likely to make an irrational decision than a participant who only invested $1 in the same auction. This realization of the sunk cost effect is moderated by the usage of an automated bidding agent. In specific, delegating actual bidding to an automated bidding agent and, consequently, incurring comparably lower *behavioral* investments significantly influences a participant's proneness to the sunk cost effect. For such participants, the likelihood of violating rationality due to sunk costs is significantly lower compared to those who place their bids manually.

This paper makes novel and significant contributions to research for several reasons. First, the literature on decision biases and IT has focused almost exclusively on DSS which *support* decision makers. Considering the increasing number of opportunities where decision makers can *delegate* their decision making to IT, it is worthwhile to investigate the impact of these new information technologies on the proneness to decision biases. Second, the impact of IT on decision biases was hitherto analyzed only in experimental settings. This is the first study that extends these results to a real market setting. Third, there are only a few studies which analyze the effects of IT usage on decision biases, and none of these analyzed the sunk cost effect. To the best of our knowledge, this paper is the first to explore this issue. Fourth, there is only limited experimental and no field evidence for the sunk cost effect for behavioral investments. We show that the delegation of decision making to IT decreases proneness to the sunk cost effect. With this result, we provide first field evidence that behavioral investments affect proneness to the sunk cost effect. In summary, we provide new insights relevant to information systems and behavioral economics research that will benefit both practitioners and researchers.

## 2. Theory and Hypotheses

### 2.1. Literature Review

Two streams of research are relevant to our study. The first examines the effect of information technologies on different decision biases and tests if and how these technologies can alleviate these biases. The second stream analyzes the sunk cost effect in different laboratory and field settings. We discuss relevant work from both of these streams in the following paragraphs.

A growing body of experimental research has examined the effect of DSS on some human decision biases. These studies, in general, have found that DSS are an effective tool to alleviate decision biases in experimental setups. In particular, earlier studies have shown that such systems can alleviate the anchoring and adjustment bias (George et al. 2000, Hoch and Schkade 1996, Lau et al. 2009), the base rate bias (Hoch and Schkade 1996, Roy and Lerch 1996), the first impression bias (Lim et al. 2000), the familiarity bias (Marett and Adams 2006), the framing bias (Bhandari et al. 2008, Cheng and Wu 2010), the representativeness bias (Bhandari et al. 2008), and the ambiguity effect (Bhandari et al. 2008). To the best of our knowledge, there exists no study, which analyzes the effect of IT usage on the sunk cost effect.

There are various approaches to alleviating decision biases. In his seminal work on debiasing, Fischhoff (1982) proposed a classification of these approaches based on the source of the decision bias. He identified faulty tasks, faulty decision makers, and mismatches between decision makers and tasks as potential sources. In case of faulty tasks, he proposed that the redesign of the task environment could debias the decision maker. When a faulty decision maker is the source of a decision bias, Fischhoff proposed an escalation design with four steps to debias the decision maker. The first step in this escalation design is to warn the decision maker about the possibility of bias. The second step is to describe the bias. The third step is to provide personalized feedback; and, the fourth is to train the decision maker extensively. As an alternative to this four-step approach, one may replace the decision maker in question. In case of a mismatch between the decision maker and the task, Fischhoff proposed to restructure the task in a way that the task and decision maker are as compatible as possible. Most of the extent studies restructured the task by modifying the presentation format (Bhandari et al. 2008, Hoch and Schkade 1996, Lau et al. 2009, Lim et al. 2000, Roy and Lerch 1996) or by providing further information (Marret and Adams 2006) and showed that this could debias the decision maker. Two other studies used warnings to debias the decision maker (Cheng and Wu 2010, George et al. 2000). We summarize this body of research in Table 1.

Our study differs from the earlier work on IT usage and decision biases in three key ways. First, we analyze a new kind of IT that can help alleviate or even prevent decision biases. Compared to DSS, automated software agents are different in terms of both functionality and purpose. The main purpose of a DSS is to *support* a decision maker, while automated software agents *act* in place of their owners. As many researchers argue that the delegation of decision making can be an effective way to alleviate decision biases (e.g. Fischhoff 1982, Roy and Lerch 1996) and since automated software agents provide numerous promising opportunities to delegate decision making, such agents are an obvious tool to alleviate or even eliminate decision biases. Nevertheless, to the best of our knowledge, none of the previous studies analyze the effect of automated software agents on decision biases. Second, we use data obtained from a real market situation. This allows us to extend the existing work on the effects of IT usage in an important way. All of the presented studies obtain their results from laboratory experiments. Therefore, it is an important issue to validate experimental results in field settings. Third, we study the impact of IT usage on the sunk cost effect. Although the sunk cost effect has important economic implications, it has not yet been analyzed in the context of IT usage and decision biases.

**Table 1: Summary of Laboratory Experiments Studying IT and Decision Biases**

|  | Bias(es) | Source(s) of Bias(es) | Debiasing Technique |
|---|---|---|---|
| Hoch and Schkade (1996) | Anchoring and adjustment, base rate bias | Faulty decision maker, mismatch between task and decision maker | Modifying the presentation format |
| Roy and Lerch (1996) | Base rate bias | Mismatch between task and decision maker | Modifying the presentation format |
| George et al. (2000) | Anchoring and adjustment | Faulty decision maker | Warning |
| Lim et al. (2000) | First impression bias | Faulty decision maker | Modifying the presentation format |
| Marett and Adams (2006) | Familiarity bias | Faulty decision maker | Providing further information |
| Bhandari et al. (2008) | Framing bias, representativeness bias, ambiguity effect | Faulty decision maker | Modifying the presentation format |
| Lau et al. (2009) | Anchoring and adjustment | Faulty decision maker | Modifying the presentation format |
| Cheng and Wu (2010) | Representativeness bias | Faulty decision maker | Warning |

The second relevant stream of literature for our research addresses the sunk cost effect. A large body of literature has analyzed this effect in laboratory and field settings. The sunk cost effect for *monetary investments* has been documented in the laboratory (e.g., Arkes and Blumer 1985, Friedman et al. 2007, Garland 1990, Staw 1976, Thaler and Johnson 1990) as well as in real market settings (e.g., Camerer and Weber 1999, Staw and Hoang 1995). With regard to purely *behavioral investments*, there is an absence of evidence from the field and only limited experimental evidence (Cunha and Caldieraro 2009, 2010, Navarro and Fantino 2009). Additionally, other scholars fail to replicate these experimental results (Otto 2010) or argue that there is no sunk cost effect for a behavioral investment, such as time (Soman 2001).

Our research differs from the prior studies on the sunk cost effect in two important ways. First, we are able to analyze the sunk cost effect for both monetary and behavioral investments in a real market setting. This gives us a unique advantage in terms of the generalizability of our results. Second, we provide not only evidence for the existence of the sunk cost effect, but also show that IT can be an effective tool for alleviating this bias. Our results might help decision makers to overcome the sunk cost effect and, thereby, substantially increase the quality of their decisions.

## 2.2. Hypotheses

As stated in the introduction, the sunk cost effect is defined as *"a greater tendency to continue an endeavor once an investment in money, time or effort has been made"* (Arkes and Blumer 1985). Thaler (1980) provides the following example for this effect: *"A family pays $40 for tickets to a basketball game to be played 60 miles from their home. On the day of the game there is a snowstorm. They decide to go anyway, but note in passing that had the tickets been given to them, they would have stayed home."* Standard economic theory predicts that, regardless of whether the family paid for the tickets or received them for free, this should not influence their decision to go to the game. Nevertheless, in this example, the family decides to attend the game because of their already sunk investments. Previous research demonstrated the sunk cost effect in several experimental and empirical settings. As is well known from this literature, higher monetary sunk costs increase the likelihood of the occurrence of the sunk cost effect (e.g., Arkes and Blumer 1985, Camerer and Weber 1999, Garland 1990, Staw and Hoang 1995). Arkes and Blumer (1985) provided funds for the Ohio University Theater to give randomly discounted season tickets to the first sixty customers. They found that the persons who got the smallest discount attended the most plays and that the number of plays attended decreased with increasing discount. In another experiment, they requested participants to decide whether they wanted to go on a skiing trip for which they had already paid $50 or on a skiing trip for which they had paid $100 but which they will enjoy less. In this experiment, the majority of the study participants decided to go on the $100 ski trip because of the higher sunk costs related to this trip. Similarly, Camerer and Weber (1999) as well as Staw and Hoang (1995) found that, after controlling for their performance, NBA players get more playing time the earlier they are picked in the draft. Being drafted ten places earlier – and thus incurring a higher sunk cost – increases the player's playing time by approximately 200 minutes over the whole season. Garland (1990) demonstrated that willingness to authorize additional resources for a threatened research and development project was both positively and linearly related to the proportion of the budget that had already been spent. Based on these findings, we arrive at the following hypothesis.

*HYPOTHESIS 1: A higher monetary investment increases proneness to the sunk cost effect.*

Research into behavioral resource allocation has documented that people recognize time costs invested in cognitive tasks (Gray et al. 2006). Accordingly, behavioral investments like investments in effort, emotional attachment to a product, or time have been cited as another source for the sunk cost effect in several studies (e.g., Cunha and Caldieraro 2009, 2010, Navarro and Fantino 2009). Cunha and Caldieraro (2009; 2010) manipulated the necessary effort and time to identify the best possible choice in a given set of alternatives. Subsequently, they added another alternative which strictly dominated the alternatives already presented. They find that participants in the high effort and time group were more likely to stick with the dominated alternative. Navarro and Fantino (2009) explore the sunk cost effect for behavioral investments using the example of time investments in four questionnaire studies and four behavioral experiments. Their results indicate that there is a consistent and robust sunk cost effect for time, if the individual has been responsible for the time investment. Based on these results, we have the following hypothesis.

*HYPOTHESIS 2: A higher behavioral investment increases proneness to the sunk cost effect.*

The delegation of decision making to prevent humans from decision biases was first proposed by Fischhoff (1982). He suggested that, as a final debiasing solution, faulty decision makers should be replaced by a superior instance. Referring to the sunk cost effect, a decision maker can be replaced in two distinct decision situations: (1) when the monetary and behavioral investments are made, or (2) when the decision maker decides whether or not to continue the endeavor. In this work, we focus on the replacement of the decision maker in the first decision situation and analyze its impact on her subsequent decisions in the latter situation. Undoubtedly, the delegation of decision making can have no influence on the nominal value of monetary investments. For instance, there is no difference in the monetary investments if somebody spends $10 on her own or if she delegates the decision to someone else, who spends $10 on her behalf. In contrast to monetary investments, the delegation of decision making can protect decision makers from behavioral investments. For example, there is a substantial difference in the behavioral investment if somebody conducts a task on her own compared to delegating this task to someone else. Thus, the delegation of decision making should protect the decision maker from potential behavioral investments. Today, rapid technological developments have enabled the delegation of decision making to IT in a large number of situations. For example, decision makers can delegate their decision making to automated bidding agents in online auctions (Adomavicius et al. 2009) or use automated trading algorithms for transacting in financial markets (Hendershott et al. 2011). These agents are widely available at a negligible marginal cost and can effectively decide on behalf of their owners. Bapna (2003) states that such automated agents "are an obvious technology that could reduce bidding costs" where these bidding costs are conceptually akin to our behavioral investments. Therefore, we have the following hypothesis.

*HYPOTHESIS 3: Delegating decision making to IT decreases behavioral investments and, thereby, reduces proneness to the sunk cost effect.*

We test our hypotheses on a unique and very rich dataset which we present in the following section.

**3. Research Setup**

*3.1. Description of the Auction Mechanism*

The data for our study come from a large German website offering pay-per-bid auctions. Between August 28, 2009 and May 9, 2010, 6,995 pay-per-bid auctions have been conducted on the website. Each such auction starts at a price of zero and with a fixed end time on a countdown clock. Auction participants are restricted to bidding a fixed bid increment (e.g., 1 cent) above the current bid and must pay a non-refundable fixed fee (e.g., 50 cents) to place each bid. Each bid extends the duration of the auction by a given time increment (e.g., 10 seconds). For example, in an auction where the current bid is $2.32 with 12 seconds on the auction countdown, an additional bid increases the current bid by 1 cent to $2.33 and extends the auction countdown by 10 seconds.[4] The participant who places the bid has to pay the fixed bidding fee of 50 cents. A participant wins the auction if her bid is not followed by another bid. The winner has to pay the current bid (in addition to the bidding fees already paid) to obtain the item. If the participant in our example is the last bidder, she would win the auctioned product for $2.33. This auction mechanism induces that the majority of bidding occurs after the initial end time[5] on the auction clock. For our website more than 98% of all bids were placed after the initial end time.

A direct buy option allows participants who do not win the auction to directly buy the auctioned item for a buy-it-now price (known prior to the commencement of the auction) net of her aggregated bidding fees paid for the bids placed in the auction. In our example, this would mean that a participant who had not won the auction but had placed 20 bids could directly buy the auctioned product for the posted buy-it-now price net of $10, which is equal to 20 times the bidding fee of 50 cents. Please note that while there can only be one winner in each auction, there is no limit to the number of participants who can buy the product directly.

We can easily transfer Thaler's (1980) example for the sunk cost effect to pay-per-bid auctions with direct buy option. *An auction participant invests $40 in bidding fees to acquire an iPhone in a pay-per-bid*

---
[4] The time increments linearly add up for each placed bid. For instance, if somebody places another bid during the same second the countdown extends by another 10 seconds to 32 seconds.

[5] The initial end time is set by the auction operator. It is the time at which an auction ends when nobody places a bid in this auction. With each placed bid, the initial end time is extended by the time increment.

*auction. After failing to win the auction, she decides to exercise the buy-it-now option, but notes in passing that had she not invested any bidding fees, she would not have bought the iPhone for the buy-it-now price net of the spent bidding fees at the auction website.*

Bid increments on the observed auction website are 0.01€ for 74%, 0.02€ for 15%, 0.05€ for 9% and 0.10€ for 2% of the auctions. The bidding fee is constant at 0.50€ for each auction while the time increment varies between 10 and 20 seconds. More than 80% of the auctions include a direct buy option for participants who fail to win the auction. After an auction is completed, each unsuccessful auction participant receives an e-mail reminding her of the opportunity to directly buy the product. The posted buy-it-now price is on average 4.8% above the price posted on amazon.de at the respective end time of the auction. The average delivery time for the direct buy items is posted in the respective auction description. It is typically ten days after receipt of money.[6]

One very important feature for our study is that participants have the opportunity to delegate their actual bidding to automated bidding agents. An agent places a predefined amount of bids autonomously. To setup such an agent, auction participants must specify the number of bids and a price interval (lower and upper limits) in which these bids should be placed. After the auction price exceeds the lower bound of the price interval, the bidding agent starts bidding. The agent places the bids at random points in time before the auction countdown is expired. It stops bidding (1) when the auction is won, (2) when the auction price exceeds the upper bound of the price interval, or (3) when the predetermined amount of bids is depleted. This bidding mechanism gives manual bidders a seemingly competitive advantage over the bidding agent. While manual bidders can wait until the very last moments of the auction, the bidding agent places its bids just at a random point in time before the auction countdown is expired. The comparison of the remaining seconds on the auction countdown shows that manual bidders place their bids on average 81 seconds before the expiration of the auction countdown. In contrast, for bidding agents this timeframe on average amounts to 310 seconds.

*3.2. Dataset*

Our dataset contains customer level bidding and transaction data for all auctions conducted between August 28, 2009 and May 9, 2010. For each auction, we have information about the actual bidding behavior of each participant, the bidding method used (agent or manual bidding), the auctioned product, whether or not the auction included a direct buy option, and if it did, whether, when, and by whom this option was exercised, and the buy-it-now price. On the participants' level, we know their date of registration, the number of bought bids as well as some demographics like gender and age. It is this level of detail that gives our field research a significant advantage over other studies of online pay-per-bid auctions (e.g., Augenblick 2010, Byers et al. 2010, Platt et. al 2012) where the data are restricted to direct observations on the websites offering such auctions. Overall, we have data for 483,414 auction participations by 87,038 distinct participants. These participants placed 6,463,642 bids in 6,995 auctions for 408 different products. Of these auctions, 5,763 included an option to directly buy the product. The direct buy option was exercised 6,337 times by 2,584 distinct participants.

---
[6] The comparably long delivery time is caused by the business model of our analyzed website. Only items that are auctioned are kept in stock. Additional items for participants who exercise the direct buy option are ordered after the receipt of money.

*3.3. Study Design*

In an ideal empirical setting, we would have data about each auction participant's private valuation of the auctioned product. This would allow us to directly observe if an auction participant executed the direct buy option due to irrationality, or just because her valuation of the product is higher than the buy-it-now price net of her spent bidding fees (we call this price the *individual direct buy price* in the following). For example, consider an auction participant who invested $40 in an auction with a buy-it-now price of $800. If this person's valuation for the product is above $760 and there is no other comparable (or better) retailer offering the product for a lower price, it is a rational decision to exercise the direct buy option after having participated in the auction. Conversely, if the valuation for the product is below $760, the direct buy decision is always attributable to irrationality. Unfortunately, it is not possible to directly observe this valuation in a non-experimental setting.

However, in line with Malmendier and Lee (2011), by utilizing publicly available price information from competing retailers, it is possible to determine a very conservative threshold for the direct buy decision scenarios that are attributable to irrationality. If the publicly available price at a comparable (or better) retailer is lower than the individual direct buy price, a direct buy decision by this participant can be attributed to irrationality. In this case, the participant's valuation for the product can be either higher or lower than the low price at the competing website. If the valuation is lower than this price, it is always irrational to buy the product. If the valuation is higher than the low price at the competing retailer, a rational auction participant would always buy the auctioned product for the lower price at the competing website.[7]

We were able to collect prices posted on amazon.de for 3,021 auctions. These prices are on average 4.8% lower than the buy-it-now price on the auction website. Given the widespread price dispersion online (Clemons et al. 2002) and the trend of higher prices being charged by reputable retailers (Smith and Brynjolfsson 2001), the posted buy-it-now price is usually much higher than the lowest price available online. Smith and Brynjolfsson (2001) as well as Clay et al. (2002) find that buyers are willing to pay a premium of approximately 10% to buy from amazon.com instead of a different, less reputable retailer with a similar offering. In addition, prior research suggests that shipment conditions and delivery times play an important role in determining a consumer's willingness to pay (Pan et al. 2002). Considering both the retailer's reputation and the delivery times, Smith and Brynjolfsson (2001) state that the majority of customers in their sample chose an offer that is 20.4% higher than the cheapest offer.

---

[7] Many bloggers and news articles suggest (e.g., Diaz 2009, Thaler 2009) that most participants of these types of auctions are deal seekers and are quite price sensitive. Accordingly, such participants are naturally quite likely to use a price comparison site to find a low price.

Given the relatively longer delivery times and the low reputation of the website we analyze, it is very unlikely that auction participants consider their individual direct buy price to be equally comparable to amazon's posted price.[8] Rather, participants are willing to pay a substantial premium to buy from amazon.de (Smith and Brynjolfsson 2001) and compare their individual direct buy prices with low prices from other retailers which are better comparable in terms of reputation and delivery times. Fortunately, the sourcing prices of the directly bought items, which the operators of the website made available to us, give us a good estimate for these low prices. As the company did not keep items in stock but directly ordered the sold items, after receiving the payment, from different online retailers, the sourcing prices are – at the minimum – representative of low prices available online. In addition, the company restricted their sourcing to a few key online retailers. Thus, there may have been even lower prices available. Consistent with the findings from prior research, the observed sourcing prices are on average 12.9% lower than respective prices posted on amazon.de. Accordingly, we identify a decision scenario as irrational if a participant's individual direct buy price is higher than the company's sourcing price of the respective product.[9]

*3.4. Main Variables*

Our research design allows us to analyze the effects of monetary and behavioral investments on irrational decisions made with regards to the direct buy option. We measure the monetary investment by a variable named *Sum of Bids* which is equal to the number of bids a participant placed in a specific auction. As participants have to pay a fixed bidding fee for each bid placed, this variable is a reliable measure for a participant's monetary investment.

---

[8] From the launch day of the website until the end of our observations period, the website sold only 6,337 units which leaves no room for building up any reputation. Additionally, the pay-per-bid auction format was met with controversy in online discussions and in the media. For example, a famous blogger called the business model *"…as close to pure, distilled evil in a business plan as I've ever seen."* These discussions might also have a negative impact on the company's reputation.

[9] In the robustness checks section, we consider alternative thresholds for such scenarios.

We identified three major sources of behavioral investments in pay-per-bid auctions: (1) the necessary investment in time and effort to participate in the auction, (2) the quasi endowment effect, and (3) the very high degree of competition in this auction format. First, there is no certainty about the end time of an auction since the duration extends with each bid placed. Therefore, participants need to monitor auctions very closely and invest a substantial amount of time and effort in their participation. Second, each time a participant places a bid, she is the highest bidder for at least a very short period of time. Even without a legal claim on the item, a bidder might develop a feeling of ownership of the item during this period (Heyman et al. 2004). This so-called 'quasi-endowment effect' further increases a participant's behavioral investment. Third, the auction mechanism in pay-per-bid auctions typically leads to a very high degree of competition among auction participants (Byers et al. 2010). In contrast to an eBay-type auction, failing to win a pay-per-bid auction not only leads to missing out on the item but also to the loss of the bidding fees already incurred. This suggests that the comparably higher level of competition in pay-per-bid auctions constitutes an additional increase in the participant's behavioral investment (for a discussion of the effect of competition on bidding behavior, see Heyman et al. 2004). Naturally, all of these sources of behavioral investments are closely related to the number of placed bids. Thus, *Sum of Bids* is also indicative of behavioral investments.

To separate the effects of behavioral investments from the effects of monetary investments, we include a variable named *Bid Agent Dummy,* as well as the interaction of *Sum of Bids* and *Bid Agent Dummy* in our model. *Bid Agent Dummy* equals one if a participant placed more than 75% of her bids in a specific auction using an automated bidding agent.[10] Holding the number of placed bids constant, the usage of an automated bidding agent cannot have any influence on a participant's monetary investment. Thus, the interaction of *Sum of Bids* and *Bid Agent Dummy* must capture any potential difference between the behavioral investment of a manual bidder and the behavioral investment of an agent owner. However, setting up an automated bidding agent may also induce some behavioral investments. These investments are captured by the variable *Bid Agent Dummy*.

To account for potential time-varying heterogeneity across auction participants, we include the variable *Number of Participations* as a historical experience measure in our model. This variable is defined as the number of participations by a specific participant in different auctions since the day of registration. Such experience measures are widely used to control for consumer heterogeneity in both marketing literature and industry practices (Anderson and Simester 2004, De et al. 2010).

---

[10] In the robustness checks section, we apply more strict definitions of agent usage.

As the economics and marketing literatures suggest, people evaluate potential savings not only in an absolute way, but also by assessing their savings in relation to the absolute price of a product (e.g., Grewal and Marmorstein 1994). For example, a $5 discount for a product with a price of $10 is valued higher than the same discount for a product worth $1,000. To control for this issue, we include a variable named *Buy-it-Now Price* in our model. It is defined as the posted buy-it-now price of each product and is measured in Euro.

'Late bidding' is a common phenomenon in online auctions like eBay (Roth and Ockenfels 2002). The auction mechanism in pay-per-bid auctions further intensifies this effect. Pay-per-bid auctions are not simply won by the bidder with the highest monetary investment, but by the bidder who placed the last bid before the auction countdown expires. Thus, it is crucial for bidders – if they do not use an automated bidding agent – to closely track the auction to the very end. Since the auction mechanism is designed in such a way that bidders have no information about the ultimate end time when they enter the auction, there might be an effect of the end time influencing the decision to directly buy a product, e.g., a bidder might evaluate her behavioral investment differently if she expected the auction to end before midnight, but continues bidding if the auction takes longer. Therefore, we divide the day into two 12 hour intervals, starting at midnight, and include one dummy variable (*Midnight - Noon Dummy*) to control for the end time of the auction. The definition of all the variables can be found in Table 2.

## 4. Empirical Analysis

### 4.1. Basic Model

The occurrence of the sunk cost effect is represented by a binary variable equaling one, if the direct buy option is exercised in an irrational decision scenario. Accordingly, we use a logistic panel regression model to examine the impact of bidding agent usage on the proneness to the sunk cost effect. In addition to some other benefits, the panel structure of our dataset allows us to control for the individual heterogeneity of auction participants (Hsiao 2003). To utilize this advantage, we can estimate both fixed and random effects models (for a general discussion of when to use fixed or random effects models, see Hsiao 2003). An important advantage of the fixed effects model is that it allows for the individual specific effects to be correlated with the explanatory variables. For the random effects model, such correlation is not allowed (Hsiao 2003). This advantage of the fixed effects model comes along with the disadvantage that this model completely ignores the between-person variations. This often yields standard errors that are considerably higher than those produced by methods considering both within- and between-person variations (Allison 2005). However, for our dataset, we expect that the individual specific effects can be correlated with the explanatory variables. For example, a person with a higher unobservable risk preference may spend more bids while participating in a pay-per-bid auction.

**Table 2: Definition of Variables**

| Name of Variable | Definition |
|---|---|
| *Direct Buy* | Dummy variable which indicates if a participant exercised the direct buy option in the respective auction. |
| *Sum of Bids* | Number of bids placed by a participant in the respective auction. |
| *Bid Agent Dummy* | Dummy variable which indicates if a participant placed more than 75% with one of the two bidding methods in a specific auction. |
| *Number of Participations* | The participant's number of auction participations before the start time of the respective auction. |
| *Buy-it-Now Price* | Price for which a participant could buy the auctioned product without having placed a bid in the respective auction. |
| *Midnight - Noon Dummy* | Dummy variable which indicates if an auction ended between midnight and noon of the next day. |
| *Number of Wins* | The participant's number of wins before the start time of the respective auction. |
| *Number of Direct Buys* | The participant's number of direct buys before the start time of the respective auction. |
| *Days since Last Participation* | The number of days since a participant's last auction participation. |
| *Sum of Bids Last Auction* | The number of bids a participant placed during her last auction participation. |
| *Gender* | The participant's gender. Takes the value one if female. |
| *Age* | The participant's age. |
| *Former Experience Bidding Agent Dummy* | Dummy variable to indicate a participant's former experience with the bidding agent. |

Accordingly, we estimate a fixed effects logistic regression model to test for a direct effect of monetary as well as behavioral sunk costs on irrational decisions to directly buy a product and, more importantly, for a potential effect of IT usage on this relationship. We include the variables *Sum of Bids* and *Bid Agent Dummy* as well as their interaction in our model. We further add the control variables presented in the preceding subsection. Thus, we consider the following model in latent variable form (Wooldridge 2010):

$$Y^*_{ij} = \alpha + \beta_1 X_{1ij} + \beta_2 X_{2ij} + \beta_3 X_{1ij} * X_{2ij} + \beta_4 D_i + \beta_5 Z_{ij} + \varepsilon_{ij}$$
$$Y_{ij} = 1\,[Y^*_{ij} > 0],\tag{1}$$

$Y_{ij}$ is a dummy variable equaling one if a participant $i$ exercises the direct buy option in auction $j$; $X_{1ij}$ denotes the sum of bids a participant placed in an auction; $X_{2ij}$ is a dummy variable indicating if a participant $i$ uses an automated bidding agent in auction $j$; $D_i$ is a set of dummy variables indicating individual fixed effects; $Z_{ij}$ is a vector of control variables; and $\varepsilon_{ij}$ is the random error term.

Note that this model specification controls for all the time invariant factors, including any differences that are inherent among participants, e.g., risk attitude, ability to use bidding agents, and intellectual capacity. More importantly, the individual fixed effects along with the time variant participant specific variable, *Number of Participations*, collectively addresses concerns regarding self-selection of participants who use an automated bidding agent. Thus, the fixed effects model allows us to address endogeneity concerns in a meaningful and robust manner (Allison 2005).

*4.2. Sample*

To isolate the impact of IT usage on the sunk cost effect, we restrict the data to instances where auction participants in a specific auction placed more than 75% of their bids either with or without an automated bidding agent. We further drop all observations where the respective auction does not include a direct buy option or where the respective participant won the auction. As the conditional fixed effects model requires variation in the independent variable (Wooldridge 2010), we restrict our sample to individuals who participated at least twice and executed the direct buy option at least once but not in each of their participations.[11] This leaves us with a sample of 230 distinct individuals who participated on average 14 times and, thereby, faced 3,153 irrational decision scenarios. In these situations the direct buy option was executed 240 times. In other words, our sample is an unbalanced panel data consisting of 230 individuals and 3,153 observations. Table 3 lists summary statistics for this sample.

*4.3. Main Results*

Our first hypothesis is supported if we find a significantly positive coefficient for the variable *Sum of Bids*. This would mean that irrational direct buy decisions do not just occur randomly; but that a participant's monetary investment has a significant positive effect on this event. This specification exactly matches the definition of the sunk cost effect. However, a positive coefficient for *Sum of Bids* does not automatically imply support for our second hypothesis as the effects of behavioral investments cannot be directly discerned from the nominal value of this variable. Hypotheses 2 and 3 are only supported if the interaction of *Sum of Bids* and *Bid Agent Dummy* has a significant negative effect on the decision to directly buy a product. A significant negative effect of this interaction term would imply that, depending on the bidding technology used, there is a significant difference between the sum of monetary and behavioral investments. As the monetary investments are the same, irrespective of which bidding method is used, different behavioral investments are the only possible explanation for this effect. This would imply both a positive effect of behavioral investments on the occurrence of irrational direct buy decisions and a negative effect of IT usage on this relationship.

---

[11] The significant test statistic (85.02) in favor of the fixed effects model for the Hausman (1978) specification test supports our model choice. Nevertheless, the results remain qualitatively unchanged if we estimate a random effects model where no such restriction is imposed.

**Table 3: Descriptive Statistics**

|  | Mean | Std. Dev. | 25th Percentile | 50th Percentile | 75 Percentile | Max |
| --- | --- | --- | --- | --- | --- | --- |
| *Direct Buy* | 0.08 | 0.27 | 0 | 0 | 0 | 1 |
| *Sum of Bids* | 13.81 | 32.19 | 1 | 3 | 11 | 501 |
| *Bid Agent Dummy* | 0.17 | 0.38 | 0 | 0 | 0 | 1 |
| *Number of Participations* | 44.95 | 77.73 | 2 | 13 | 50 | 444 |
| *Buy-it-Now Price* (in €) | 364.22 | 319.47 | 129.97 | 236.58 | 579.00 | 1,336.99 |
| *Midnight - Noon Dummy* | 0.28 | 0.45 | 0 | 0 | 1 | 1 |

Table 4 presents the estimates of the fixed effect model. Confirming our first hypothesis, we find a very strong and highly significant positive relationship between the number of bids placed and the likelihood to exercise the direct buy option. Consistent with hypothesis 2 and hypothesis 3, the coefficient for the interaction of *Sum of Bids* and *Bid Agent Dummy* is negative and highly significant.

### *4.4. Economic Significance of Main Results*

As we estimate a logistic regression model, the coefficients cannot be interpreted as the change in the mean of $Y_{ij}$ for a one unit increase in the respective predictor variable, with all other predictors remaining constant.[12] Rather, they can be interpreted as the natural logarithm of a multiplying factor by which the predicted odds of $Y_{ij} = 1$ change, given a one unit increase in the predictor variable, holding all other

---

[12] This applies especially to interaction effects in logit (and other non-linear) models. In such models, the magnitude of the interaction effect does not equal the marginal effect of the interaction term and can even be of opposite sign (Ai and Norton 2003). However, it is unproblematic to interpret these interaction effects using multiplicative effects like odds ratios (Buis 2010).

**Table 4: Main Results**

| Variable | |
|---|---|
| Sum of Bids | 0.0937*** |
|  | (0.0078) |
| Bid Agent Dummy | 0.9883*** |
|  | (0.3773) |
| Sum of Bids * Bid Agent Dummy | -0.0468*** |
|  | (0.0071) |
| Number of Participations | 0.0114* |
|  | (0.0065) |
| Buy-it-Now Price | -0.0075*** |
|  | (0.0009) |
| Midnight – Noon Dummy | 0.2144 |
|  | (0.2554) |
| Log likelihood | -192.52 |
| Number of observations | 3,153 |
| Number of participants | 230 |

Note: Standard errors are in parentheses.
* $p < 0.10$; ** $p < 0.05$; *** $p < 0.01$.

predictor variables constant.[13] Therefore, we first have to calculate the odds ratio, which is equal to the exponent of the coefficient of the respective variable.

Table 4 shows that the coefficient associated with *Sum of Bids* is 0.0937. Thus the odds ratio for this variable is equal to exp(0.0937) = 1.0982. Because *Sum of Bids* is part of the interaction term (*Sum of Bids * Bid Agent Dummy*), the coefficient does not represent a main effect but instead represents a conditional effect, i.e., the effect of a one unit increase of *Sum of Bids* when the moderator variable is zero (Jaccard 2001). Thus, 1.0982 is the multiplying factor by which the odds of buying the product directly changes for each additionally placed bid for participants who do not use an automated bidding agent. To put it differently, the odds for buying the product directly increases by about 10% for each additional manually placed bid.

Next, we want to assess the effect of *Sum of Bids* on the odds of buying the product directly when a participant uses an automated bidding agent. Here, we need to take the exponent of the sum of the coefficients of *Sum of Bids* and the interaction term (*Sum of Bids * Bid Agent Dummy)*. The resulting odds ratio is 1.0480, implying that a one unit increase in *Sum of Bids* increases the odds of directly buying the product by 5%. This is only 50% of the increase for participants who do not use an automated bidding

---

[13] The odds are defined as $\text{Odds} = \frac{P(Y_{ij}=1)}{\left(1-P(Y_{ij}=1)\right)}$.

**Figure 1: Odds Ratio as a Function of *Sum of Bids***

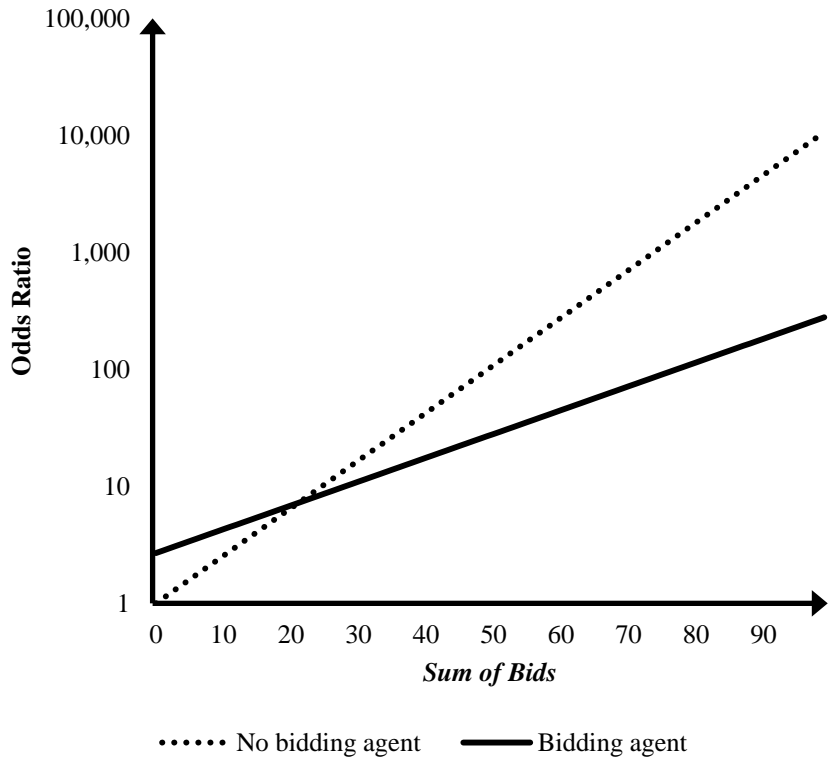

⋯⋯ No bidding agent     —— Bidding agent agent. For example, consider a participant who placed 67 bids in an auction.[14] If this participant had placed her bids manually, the odds of buying the product directly are approximately 533 times higher than if she had not placed any bid. Conversely, if this participant has placed her bids via an automated bidding agent, her odds are only 23 times higher than if she had not placed any bid. This result strongly supports our hypotheses 2 and 3.

In addition to the indirect effect, there is also a direct effect of *Bid Agent Dummy* on the proneness to the sunk cost effect. To assess this effect, we need to take the exponent of the coefficient of *Bid Agent Dummy*. The resulting odds ratio is 2.6867. This odds ratio implies that the base odds of buying the product directly must be multiplied by 2.6867 if a participant switches from manual bidding to using an automated bidding agent.

To analyze the total effect of bidding agent usage on the likelihood to exercise the direct buy option, we need to combine the effects of *Sum of Bids* and *Bid Agent Dummy* for participants who use an automated bidding agent and compare this combined effect with the effect of *Sum of Bids* for manual bidders.

---

[14] This is equal to the average number of bids placed by a participant who exercised the option to directly buy the product.

**Table 5: Summary of Results**

| Hypothesis | Result |
|---|---|
| H1: *A higher monetary investment increases proneness to the sunk cost effect.* | Supported |
| H2: *A higher behavioral investment increases proneness to the sunk cost effect.* | Supported |
| H3: *IT usage decreases behavioral investments and, thereby, reduces proneness to the sunk cost effect.* | Supported |

Figure 1 shows the resulting odds ratios for participants who use an automated bidding agent and for manual bidders.[15] We see a *positive* effect of bidding agent usage on the likelihood to exercise the direct buy option for the first 21 bids. If we continue the example from the last paragraph and consider a participant who placed 67 bids using an automated bidding agent, her odds are 62 times higher than if she had not placed any bid. If this participant placed her bids manually, her odds are still 533 times higher than if she had not placed any bid.

The positive effect for the first 21 bids can be attributed to the average fixed behavioral investments necessary to set up an automated bidding agent. This finding gives us additional support for our second hypothesis but is seemingly contrary to our third hypothesis. But a closer look reveals that the average predicted probability of exercising the direct buy option for participants who placed less than 22 bids with an automated bidding agent is 1.3%. Therefore, the positive effect of bidding agent usage on the likelihood to exercise the direct buy option is negligible. In contrast, for the participants who placed more than 21 bids with an automated bidding agent, the total effect of bidding agent usage is always negative as stated by our third hypothesis. Table 5 summarizes these results.

Our research design allows us to quantify the difference of the behavioral investments of agent users and manual bidders in monetary value. For each number of manually placed bids, we can easily find the respective amount of bids an agent user needs to place to have a similar odds ratio. The difference between these two numbers multiplied by the bidding fee is equal to the monetary value of the disparity of the behavioral investment if the respective participant made use of an automated bidding agent. Figure 2 shows this difference (solid line) as well as the ratio of this difference and a participant's total investment (dashed line) as a function of the number of placed bids. We see from Figure 2 that the behavioral investment of a manual bidder who placed 67 bids (which is equal to a monetary investment of 33.5€) has a monetary value of 22.5€. So, for such a bidder, the behavioral investments sum up to

---
[15] Note that the y-axis has a logarithmic scale.

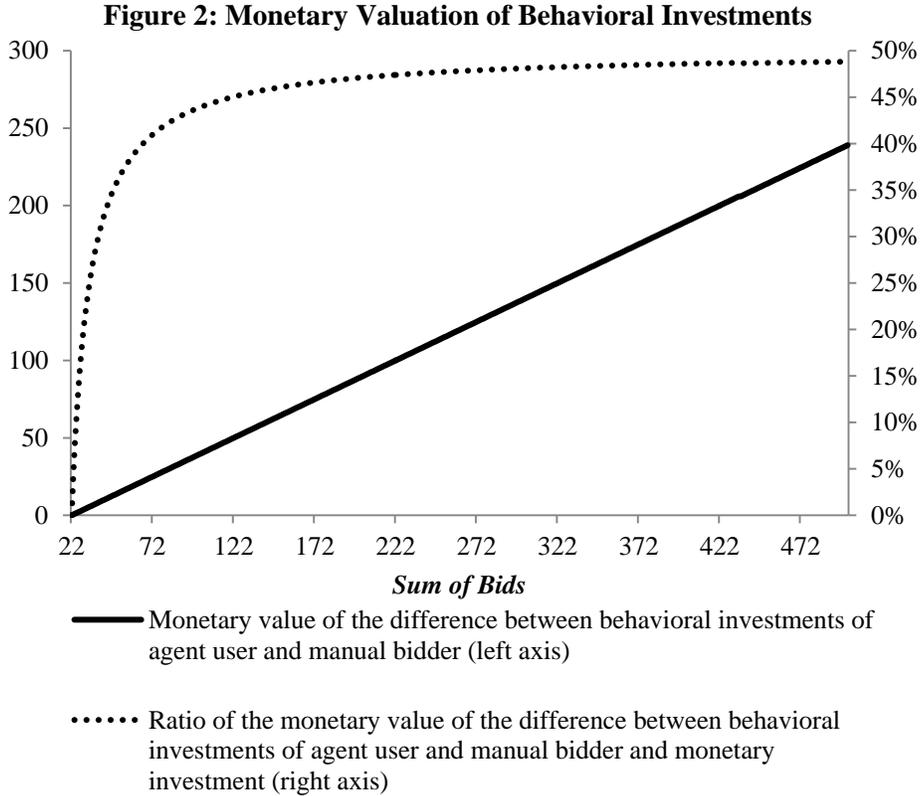

Figure 2: Monetary Valuation of Behavioral Investments

— Monetary value of the difference between behavioral investments of agent user and manual bidder (left axis)

······ Ratio of the monetary value of the difference between behavioral investments of agent user and manual bidder and monetary investment (right axis) approximately 40% of her total investments. With increasing number of placed bids the ratio between the monetary value of the behavioral investment and the total investment approaches to 50%.

## 5. Robustness Checks

Although our main results confirm our hypotheses, we have examined a number of competing explanations for the observed effects. In the following, we demonstrate that our results survive a wide range of robustness checks.

### *5.1. Quality of External Price*

One may argue that the company's sourcing price is not a perfect estimate for the reference price of auction participants. Even though the website we analyze has no reputation and comparably long delivery times, there may be participants who have a preference for buying at our auction website. Therefore, these participants do not compare their individual direct buy price with a low price available online, but with a price from – in their view – a more comparable retailer. In addition, even if it is an empirically well-founded assumption, we do not exactly know if the customers are aware of the low price. If the customers are only aware of prices higher than the companies sourcing price or if the customers are willing to pay a premium to buy at our analyzed website, we would potentially misidentify some rational direct buy decisions as irrational. Furthermore, there may be also some costs associated with switching from our auction website to a retailer offering the low price. If these switching costs are higher than the difference between the low price and the individual direct buy price, the respective direct buy decisions cannot be attributed to irrationality. We can address these concerns by increasing our estimate for a low price available online. Accordingly, we re-estimate our main model on datasets where we stepwise raise the threshold where we identify a decision scenario as irrational. In the first step, we look only at participants who face decision scenarios where their individual direct buy price was above 105% of the companies sourcing price. Subsequently, we increase this threshold value to 110% of the sourcing price. As a very strict robustness check, we compare a participant's individual direct buy price with the price posted on amazon.de at the end time of the auction.[16] For this robustness check, we identify only those decision scenarios as irrational where the individual direct buy price was higher than the respective price posted on amazon.de. This is a very strict robustness check as it assumes that participants have no additional willingness to pay to buy at a reputable retailer or to get a shorter delivery time. These assumptions are in direct contradiction to the empirical results of prior research (e.g., Clay et al. 2002, Smith and Brynjolfsson 2001). We present the results of these robustness checks in columns one to three of Table 6. Reassuringly, our main coefficients remain qualitatively unchanged. These results reaffirm our key findings and show that the results from our main model are not biased due to a wrong estimate of a low price available online.

## 5.2. Self-Selection Arising from Bidding Agent Usage

As mentioned before, the fixed effects model can be quite effective in addressing potential self-selection issues; still, we can further verify the robustness of our results by utilizing the propensity score matching method suggested by Rosenbaum and Rubin (1983).[17] This method is commonly used to control for self-selection effects due to individual specific time variant and time invariant unobserved heterogeneity. For example, this approach considers the impact of familiarity or experience with the auction platform on bidding agent usage and the sunk cost effect – that is controlled for in the fixed effects model through the variable *Number of Participations* – in a different way. Its basic idea is to find a sample of the control group (i.e., participants who do not use the automated bidding agent) that is similar to the treatment group (i.e., participants who use an automated bidding agent) in all relevant pretreatment characteristics. When

---

[16] Note that this approach is equivalent to considering switching costs of 5%, 10% and 12.9% of the buy-it-now price.

[17] We have used the STATA PSMATCH2 module by Leuven and Sianesi (2009) to implement propensity score matching.

**Table 6: Results with Consideration of Higher External Prices**

| Variable | 105% Sourcing Price | 110% Sourcing Price | Amazon Price |
|---|---|---|---|
| Sum of Bids | 0.0966*** | 0.1004*** | 0.1693*** |
|  | (0.0097) | (0.0126) | (0.0462) |
| Bid Agent Dummy | 1.2794*** | 1.3758*** | -0.7218 |
|  | (0.4315) | (0.4917) | (1.1421) |
| Sum of Bids * Bid Agent Dummy | -0.0590*** | -0.0677*** | -0.0706* |
|  | (0.0095) | (0.0129) | (0.0366) |
| Number of Participations | 0.0293** | 0.0238* | -0.0057 |
|  | (0.0125) | (0.0138) | (0.0254) |
| Buy-it-Now Price | -0.0058*** | -0.0049*** | -0.0100** |
|  | (0.0009) | (0.0009) | (0.0048) |
| Midnight – Noon Dummy | -0.0234 | 0.1969 | -1.1495 |
|  | (0.2916) | (0.3224) | (1.1241) |
| Log likelihood | -142.14 | -111.15 | -21.07 |
| Number of observations | 1,770 | 1,186 | 236 |
| Number of participants | 165 | 119 | 50 |

Note: Standard errors are in parentheses.
* $p < 0.10$; ** $p < 0.05$; *** $p < 0.01$.

this is done, the difference in outcomes for both groups can be attributed solely to the treatment rather than to self-selection effects.

We use the variable *Bid Agent Dummy* to indicate whether an observation belongs to the treatment or to the control group. Then, in order to calculate the propensity score for each observation, we estimate a logit model with *Bid Agent Dummy* as dependent variable and variables measuring participants' experience with the auction website (*Number of Participations*, *Number of Wins*, and *Number of Direct Buys, Days since Last Participation*), participants' behavior in the last auction (*Sum of Bids Last Auction*), participants' *Gender* and *Age*, and participants' technology affinity (*Former Experience Bidding Agent Dummy*).[18] Subsequently, to create an appropriate control group, we use caliper matching without replacement to match each treated observation to its nearest untreated neighbor based on the propensity score. The violation of the common support condition is a major source of evaluation bias in matching studies (Heckman et al. 1997). Accordingly, we restrict the maximal difference of the propensity scores of a treated and an untreated individual to 0.01. This approach ensures that we only match observations where the common support condition is not violated (Caliendo and Kopeinig 2008). We find that before matching, the control group is significantly different from the treatment group on dimensions such as experience with the auction website or technology affinity. However, after matching,

---
[18] Definitions for all of these variables can be found in Table 2.

**Table 7: Results for Propensity Score Matching and Product Category Dummies**

| Variable | Propensity Score Matching | Product Category Dummies |
|---|---|---|
| Sum of Bids | 0.0971*** | 0.0928*** |
|  | (0.0147) | (0.0078) |
| Bid Agent Dummy | 1.563*** | 0.9585** |
|  | (0.496) | (0.3775) |
| Sum of Bids * Bid Agent Dummy | -0.0514*** | -0.0461*** |
|  | (0.0146) | (0.0072) |
| Number of Participations | -0.0193*** | 0.0109 |
|  | (0.00455) | (0.0067) |
| Buy-it-Now Price | -0.0064*** | -0.0073*** |
|  | (0.0011) | (0.0009) |
| Midnight – Noon Dummy | 0.249 | 0.2494 |
|  | (0.326) | (0.2587) |
| Computers Dummy |  | -0.5414 |
|  |  | (0.4284) |
| Others Dummy |  | 0.0464 |
|  |  | (0.3194) |
| Videogames Dummy |  | 0.2725 |
|  |  | (0.2945) |
| Intercept | -2.884*** |  |
|  | (0.472) |  |
| Log likelihood | -158.68 | -190.94 |
| Number of observations | 1,028 | 3,153 |
| Number of participants | 173 | 230 |

Note: Standard errors are in parentheses.
* $p < 0.10$; ** $p < 0.05$; *** $p < 0.01$.

the difference between the matched control group and the treatment group is insignificant on all the dimensions.

As the treatment and control groups do not differ significantly in all pre-treatment characteristics, we can use both, within- and between-person variation to estimate our model for this robustness check. Accordingly, we estimate our model in random effects specification for this matched sample. The results presented in the first column of Table 7 are very similar to those in Table 4. As in our main analysis, the coefficient for the variable *Sum of Bids* is positive and highly significant while the coefficient for the interaction term of *Sum of Bids* and *Bid Agent Dummy* is negative and highly significant. These results reinforce our confidence in our results not being biased through the self-selection of participants using an automated bidding agent.

### 5.3. Considering Product Category Specific Effects

One may argue that there are product category specific effects that influence our results. For example, in a recent study of pay-per-bid auctions, Platt et al. (2012) find deviating bidding behavior for products from the category videogames in their dataset. These effects may not be completely captured by the variable *Buy-it-Now Price*. We address this issue by assigning each product to one of four product categories (home electronics, computers, others, and videogames) and include dummy variables for three of these categories in our model (*Computers Dummy*, *Others Dummy*, *Videogames Dummy*). The second column in Table 7 shows that our results remain qualitatively unchanged for this robustness check.

*5.4 Nonparametric Matching Model*

An additional way to rule out alternative explanations, as well as to further assess the robustness of our main results, is to follow Pope and Schweitzer (2011) and conduct nonparametric analyses. Although we lose some statistical power with this approach, it enables us to compare agent users and manual bidders in a novel way. Like Pope and Schweitzer (2011), we consider a matching model to compare participations with and without automated bidding agent usage where participants placed the same number of bids for the same product. As a first step, we create a list of irrational decision scenarios where the respective participant placed her bids using an automated bidding agent. Next, we use a matching algorithm to identify irrational decision scenarios where the respective participant placed the same number of bids for exactly the same product manually. For some scenarios we do not find any exact match, while for other irrational decision scenarios we find more than one exact match. In the first case, we do not include this scenario into our analysis. In the latter case, we include each matched pair into our analysis.[19]

We report results from our matched analysis in Table 8. The first column shows the 238,569 pairs of irrational decision scenarios with and without automated bidding agent where we could find an exact match for the sum of placed bids for the same product across auctions. Consistent with our hypotheses and our parametric results, for the same number of placed bids for the same product, the direct buy option is executed significantly more often in irrational decision scenarios where bids are placed manually (0.0541% of the participations versus 0.0117% of the participations, $p < 0.001$). In columns two and three, we report results from pairs of irrational decision scenarios with and without automated bidding agent where the number of placed bids differ by 5% and 10%, respectively. With larger differences in the number of placed bids, we increase the count of matched pairs, but of course, these matches are less precise. We find that manual bidders exercise the direct buy option four and three times, respectively, more often than participants who use an automated bidding agent. Imposing an additional restriction that we only match decision scenarios where the respective participant placed the same number of bids for the

---

[19] Consider the following example: We identify an irrational decision scenario where a participant places 10 bids for a specific product with an automated bidding agent. For the same product across different auctions, we find 5 irrational decision scenarios where participants also place 10 bids manually. This results in 5 matched pairs (the automated bidding agent scenario matched to each of the manually bidding scenarios).

**Table 8: Nonparametric Matched Sample Analysis**

|  | Maximum difference in the number of placed bids for the same product between matched irrational decision scenarios with and without bidding agent *across auctions* | | | Maximum difference in the number of placed bids for the same product *in the same auction* between matched irrational decision scenarios with and without bidding agent | | |
| --- | --- | --- | --- | --- | --- | --- |
|  | Exact match | ≤ 5% | ≤ 10% | Exact match | ≤ 5% | ≤ 10% |
| Fraction of executed direct buy options (automated bidding agent) | 0.0117% (1.0833%) | 0.0053% (0.7280%) | 0.0096% (0.9778%) | 0.0031% (0.5601%) | 0.0045% (0.6672%) | 0.0106% (1.0275%) |
| Fraction of executed direct buy options (manual bidding) | 0.0541% (2.3247%) | 0.0221% (1.4876%) | 0.0282% (1.6802%) | 0.0230% (1.5166%) | 0.0298% (1.7268%) | 0.0392% (1.9826%) |
| Average number of placed bids (automated bidding agent) | 4.8359 (8.9542) | 5.2359 (7.4805) | 6.8607 (9.2563) | 4.8898 (5.9406) | 5.8543 (8.4688) | 7.6568 (10.1796) |
| Average number of placed bids (manual bidding) | 4.8359 (8.9542) | 5.2319 (7.4540) | 6.8134 (9.1907) | 4.8898 (5.9406) | 5.8504 (8.4607) | 7.5980 (10.1168) |
| Number of pairs | 238,569 | 6,829,613 | 8,095,428 | 95,630 | 224,633 | 274,653 |

Note: Standard errors are in parentheses.

same product in the *same auction*, we emulate the first column in column four of Table 8. Applying the same restriction, we emulate columns two and three in columns five and six of Table 8. Reassuringly, we find that manual bidders exercise the direct buy option between four and seven times more often than participants who use an automated bidding agent. Overall, the nonparametric results qualitatively echo the results obtained from the main model.

*5.5. Additional Robustness Checks*

We conduct a wide range of further robustness checks and find qualitatively similar results.[20] First, there might be concerns that some participants do not have enough knowledge of the pay-per-bid auction mechanism when participating for the first time. In particular, some participants may not really understand the mechanism of the direct buy option, although all relevant information is publicly available before the auction starts. To address this issue, we drop each observation with no prior auction participation from our dataset and re-estimate our regression models on the resulting subsample. Second, we use more precise controls for the end time of the auction. We split the day into four six hour intervals and include three dummy variables controlling for the end time of an auction into our model. Third, we use more strict definitions for the usage of an automated bidding agent. In the first step we only include observations where more than 90% of the bids were placed with one of the two bidding methods. In the second step, we increase this threshold to 100%. Fourth, there might be additional concerns that buyers might know about the lower price, but stick with our website because of some potential switching costs. One may argue that these costs are not a percentage of the product value, but a fixed nominal amount. We address this issue by including only those observations in our sample where the difference between the price on our website and the reference price was above 5€. Finally, all of our results are robust to random effects specifications (probit and logit) and linear probability model specifications (random and fixed effects) as well.

---

[20] The results of these robustness checks are available upon request.

## 6. Discussion and Conclusion

The advent of electronic markets has largely increased people's opportunities to delegate their decision making to IT. Intelligent bidding agents on auction platforms (Adomavicius et al. 2009) or algorithms making their own decisions when trading on financial markets (Hendershott et al. 2011) are only a few examples for these opportunities. It is surprising, then, that there has not been any empirical research to date on how this delegation might affect human decision making and, especially, how it might influence critical human decision biases. This paper attempts to fill this gap in the literature. Our analysis shows that participants' IT usage has a significant impact on their proneness to a critical human decision bias – the sunk cost effect. More specifically, participants who delegate parts of their decisions to an automated bidding agent are far less prone to this bias. In economic terms, delegation to IT reduces the impact of overall investments on the sunk cost effect by 50%. This result can be attributed to the comparably lower behavioral investments for participants who place their bids using an automated bidding agent.

With this finding, we contribute also to the literature on the sunk cost effect. We are the first to observe both monetary and behavioral investments in a real market setting and to analyze their respective and relative roles on the sunk cost effect. Our findings support the lab experiments of Cunha and Caldieraro (2009, 2010) who stated that even purely behavioral sunk costs can induce the sunk cost effect. In addition, our findings could also help explain why people often overbid in online auctions (Malmendier and Lee 2011). Ariely and Simonson (2003) found that participants in 494 out of 500 eBay-type online auctions paid a higher price for the auctioned item than the price found on some online retailers' websites. One explanation for this observation could be that auction participants who invested a substantial amount of time and effort in their participation are overbidding because they prefer to win the auctioned item for a slightly higher price compared with realizing the loss of all of their behavioral investments.

Most importantly, our findings are a first step towards generalizing the earlier works on the effect of IT usage on decision biases to real market situations. Our results show that IT can alleviate the sunk cost effect in a real market situation. However, our work is just a first step in this generalization process. Further research is needed to examine if IT can also alleviate or even eliminate other decision biases in real market situations.

The findings in this paper are robust and have survived a wide range of robustness checks. The data analyzed in our paper were provided directly by a website offering pay-per-bid auctions. Therefore, we are able to include a wide range of controls into our models. Additionally, our results remain qualitatively unchanged when we consider different reference prices for the auction participants. We also consider the possibility that participants who use an automated bidding agent are self-selected and, therefore, differ from those who do not use an automated bidding agent, reassuring the robustness of our results.

Although we find strong evidence for the effects described above, we cannot apply our findings from pay-per-bid auctions directly to other domains. Nevertheless, our results are suggestive. Since the delegation of decision making to IT has a statistically and economically significant impact on the occurrence of the sunk cost effect, behavioral sunk costs may have a much higher impact than prior research suggests (Soman 2001). Further research, particularly experimental studies that randomly manipulate participants' delegation to IT would be able to present additional evidence for this effect on other markets.

Understanding the impact of delegating parts of the decision making to IT on the sunk cost effect has important managerial implications. Our study shows that not only monetary investments but also behavioral investments affect the occurrence of the sunk cost effect. These behavioral investments can occur in a wide range of situations, e.g., project management, policy making, financial markets, and auctions. Therefore, decision makers and, especially, managers should analyze all types of decision situations very carefully for potential behavioral investments. In case of substantial behavioral investments, the introduction of a software agent or another dispassionate advisor can detach decision makers from the decision process and, thus, prevent them from behavioral investments. This detachment could substantially increase the quality of their decision processes in various situations and, hence, provide them with a competitive advantage.

## 7. References

Adomavicius, G., A. Gupta, D. Zhdanov. 2009. Designing intelligent software agents for auctions with limited information feedback. *Information Systems Research* 20(4) 507-526.

Ai, C., E. C. Norton. 2003. Interaction terms in logit and probit models. *Economic Letters* 80(1), 123-129.

Allison, P. D. 2005. *Fixed effects regression methods for longitudinal data*. SAS Institute, Cary, NC.

Anderson, E., D. Simester. 2004. Long-run effects of promotion depth on new versus established customers: Three field studies. *Marketing Science* 23(1), 4-20.

Arkes, H. R., C. Blumer. 1985. The psychology of sunk cost. *Organizational Behavior and Human Decision Processes* 35(1), 124-140.

Ariely, D., I. Simonson. 2003. Buying, bidding, playing or competing? Value assessment and decision dynamics in online auctions. *Journal of Consumer Psychology* 13(1), 113-123.

Augenblick, N. 2010. Consumer and producer behavior in the market for penny auctions: A theoretical and empirical analysis. Unpublished manuscript. Available at: http://faculty.haas.berkeley.edu/ned/index.html.

Bapna, R. 2003. When snipers become predators: Can mechanism design save online auctions? *Communications of the ACM* 46(12), 152-158.

Bhandari, G., K. Hassanein, R. Deaves. 2008. Debiasing investors with decision support system: An experimental investigation. *Decision Support Systems* 46(1) 399-410.

Buis, M. 2010. Interpretation of interactions in non-linear models. *The Stata Journal* 10(2), 305-308.

Byers, J. W., M. Mitzenmacher, M., G. Zervas. 2010. Information asymmetries in pay-per-bid auctions. *Proceedings of 11th ACM Conference on Electronic Commerce*, Cambridge MA, June 2010, 1-12.

Caliendo, M., S. Kopeinig. 2008. Some practical guidance for the implementation of propensity score matching. *Journal of Economic Surveys* 22(1), 31-72.

Camerer, C. F., R. A. Weber. 1999. The econometrics and behavioral economics of escalation of commitment: A re-examination of Staw and Hoang's NBA data. *Journal of Economic Behavior and Organization* 39(1), 59-82.

Camerer, C. F., G. Loewenstein, M. Rabin. 2004. *Advances in behavioral economics*. Princeton University Press, Princeton, NY.

Chang, K. 2002. *New age bidding – are bots smarter?* Available at: http://www.pricingsociety.com/Art_New_Age_Bidding_Are_Bots_Smarter_FF.htm.

Cheng, F., C. Wu. 2010. Debiasing the framing effect: The effect of warning and involvement. *Decision Support Systems* 49(3), 328-334.

Clay, K., R. Krishnan, E. Wolff, D. Fernandes. 2002. Retail Strategies on the Web: Price and Non-Price Competition in the Online Book Industry. *Journal of Industrial Economics* 50(3), 351-367.

Clemons, E. K., I. Hann, L. M. Hitt. 2002. Price dispersion and differentiation in online travel: An empirical investigation. *Management Science* 48(4), 534-549.

Cunha, M. Jr., F. Caldieraro. 2009. Sunk-cost effects on purely behavioral investments. *Cognitive Science* 33(1), 105-113.

Cunha, M. Jr., F. Caldieraro. 2010. On the observability of purely behavioral sunk-cost effects: Theoretical and empirical support for the BISC model. *Cognitive Science* 34(8), 1384-1387.

De, P., Y. Hu, M. S. Rahman. 2010. Technology usage and online sales: An empirical study. *Management Science* 56(11), 1930-1945.

DellaVigna, S. 2009. Psychology and economics: Evidence from the field. *Journal of Economic Literature* 47(2), 315-372.

Diaz, S. 2009. A $79.16 MacBook? On Swoopo, if it sounds too good to be true... ZDNet, March 9, http://www.zdnet.com/blog/btl/a-7916-macbook-on-swoopo-if-it-sounds-too-good-to-be-true/14122.

Fischhoff, B. 1982. Debiasing. In D. Kahnemann, P. Slovic, A. Tversky (Eds.) Judgement Under Uncertainty: Heuristics and Biases. Cambridge University Press, New York.

Friedman, D., K. Pommerenke, R. Lukose, G. Milam, B. A. Huberman. 2007. Searching for the sunk cost fallacy. *Experimental Economics* 10(1), 79-104.

Garland, H. 1990. Throwing good money after bad: The effect of sunk costs on the decision to escalate commitment to an ongoing project. *Journal of Applied Psychology* 75(6), 728-731.

George, J. F., K. Duffy, M. Ahuja. 2000. Countering the anchoring and adjustment bias with decision support systems. *Decision Support Systems* 29(2), 195-206.

Gray, W. D., C. R. Sims, W. Fu, M. J. Schoelles. 2006. The soft constraints hypothesis: A rational analysis approach to resource allocation for interactive behavior. *Psychological Review* 113(3), 461-482.

Greenwald, A., J. Boyan. 2001. Bidding algorithms for simultaneous auctions. *Proceedings of 3rd ACM Conference on Electronic Commerce*, New York NY, October 2001, 115-124.

Grewal, D., H. Marmorstein. 1994. Market price variation, perceived price variation and consumers' price search decisions for durable goods. *Journal of Consumer Research* 21(3), 453-460.

Guo, Z., G. J. Koehler, A. B. Whinston. 2011. A computational analysis of bundle trading markets design for distributed resource allocation. *Information Systems Research* Forthcoming.

Harford, T. 2010. The auction site that's pure temptation. *The Undercover Economist* (March 13), available at: http://timharford.com/2010/03/the-auction-site-that's-pure-temptation.

Hausman, J. A. 1978. Specification tests in econometrics. *Econometrica* 46(6), 1251-1271.

Heckman, J. J., H. Ichimura, P. Todd. 1997. Matching as an econometric evaluation estimator: Evidence from evaluating a job training programme. *Review of Economic Studies* 64(4), 605-654.

Hendershott, T., C. M. Jones, A. J. Menkveld. 2011. Does algorithmic trading improve liquidity? *Journal of Finance* 66(1), 1-33.

Heyman, J. E., Y. Orhun, D. Ariely. 2004. Auction fever: The effect of opponents and quasi-endowment on product valuations. *Journal of Interactive Marketing* 18(4), 7-21.

Hinz, O., I. Hann, M. Spann. 2011. Price discrimination in e-commerce? An examination of dynamic pricing in name-your-own-price markets. *MIS Quarterly* 35(1), 81-98.

Hoch, S. J., D. A. Schkade. 1996. A psychological approach to decision support systems. *Management Science* 42(1), 51-64.

Hsiao, C. 2003. *Analysis of panel data*. Cambridge University Press, Cambridge, MA.

Jaccard, J. 2001. *Interaction effects in logistic regression*. Sage, Thousand Oaks, CA.

Kahneman, D., A. Tversky. 1979. Prospect theory: An analysis of decision under risk. *Econometrica* 47(2), 263-292.

Kanodia, C., R. Bushman, J. Dickhaut. 1989. Escalation errors and the sunk cost effect: An explanation based on reputation and information asymmetries. *Journal of Accounting Research* 27(1), 59-77.

Lau, A. Y. S., E. W. Coiera. 2009. Can cognitive biases during consumer health information searches be reduced to improve decision making? *Journal of the American Medical Informatics Association* 16(1), 54-65.

Leuven, E., B. Sianesi. 2009. PSMATCH2: Stata module to perform full Mahalanobis and propensity score matching, common support graphing, and covariate imbalance testing. Statistical software component S432001. Boston College Department of Economics, Chestnut Hill, MA.

Levitt, S. D., J. A. List. 2008. Homo economicus evolves. *Science* 319(5865), 909-910.

Lim, K. H., I. Benbasat, L. M. Ward. 2000. The role of multimedia in changing first impression bias. *Information Systems Research* 11(2), 115-136.

List, J. A. 2003. Does market experience eliminate market anomalies? *Quarterly Journal of Economics* 118(1), 41-71.

Mackenzie, M. 2009. High-frequency trading under scrutiny, *Financial Times* (July 28), available at: http://www.ft.com/intl/cms/s/0/d5fa0660-7b95-11de-9772-00144feabdc0.html#axzz1Xi3gh3ID.

Malmendier, U., Y. H. Lee. 2011. The bidder's curse. *American Economic Review* 101(2), 749-787.

Marret, K., G. Adams. 2006. The role of decision support in alleviating the familiarity bias. *Proceedings of the 39th Hawaii International Conference on System Sciences* 2, 31b.

Navarro, A. D., E. Fantino. 2009. The sunk time effect: An exploration. *Journal of Behavioral Decision Making* 22(3), 252-270.

Nisbett, R., L. Ross. 1980. *Human inference: Strategies and shortcomings of social judgement*. Prentice-Hall, Englewood Cliffs, NJ.

Otto, A. R. 2010. Three attempts to replicate the behavioral sunk-cost effect: A note on Cunha and Caldieraro (2009). *Cognitive Science* 34(8), 1379-1383.

Pan, X., B. T. Ratchford, V. Shankar. 2002. Can price dispersion in online markets be explained by differences in e-tailer service quality? *Journal of the Academy of Marketing Sciences* 30(4), 433-445.

Platt, B. C., J. Price, H. Tappen. 2012. Pay-to-bid auctions. Unpublished manuscript. Available at: http://economics.byu.edu/SiteAssets/Pages/Faculty/Brennan%20Platt/PayToBid.pdf.

Pope, D. G., M. E. Schweitzer. 2011. Is Tiger Woods loss averse? Persistent bias in the face of experience, competition and high stakes. *American Economic Review* 101(1), 129-157.

Rosenbaum, P., D. Rubin. 1983. The central role of the propensity score in observational studies for causal effects. *Biometrika* 70(1), 41-55.

Roth, A. E., A. Ockenfels. 2002. Last-minute bidding and the rules for ending second-price auctions: Evidence from eBay and Amazon auctions on the internet. *American Economic Review* 92(4), 1093-1103.

Roy, M. C., F. J. Lerch. 1996. Overcoming ineffective mental representations in base-rate problems. *Information Systems Research* 7(2), 233-247.

Smith, M. D., E. Brynjolfsson. 2001. Consumer decision-making at an internet shop bot: Brand still matters. *Journal of Industrial Economics* 49(4), 541-558.

Soman, D. 2001. The mental accounting of sunk time costs: Why time is not like money. *Journal of Behavioral Decision Making* 14(3), 169-185.

Staw, B. M. 1976. Knee deep in the big muddy: A study of escalating commitment to a chosen course of action. *Organization Behavior and Human Performance* 16(1), 27-44.

Staw, B. M., H. Hoang. 1995. Sunk costs in the NBA: Why draft order affects playing time and survival in professional basketball. *Administrative Science Quarterly* 40(3), 474-494.

Stone, P., A. Greenwald. 2005. The first international trading agent competition: Autonomous bidding agents. *Electronic Commerce Research* 5(2), 229-265.

Thaler, R. H. 1980. Toward a positive theory of consumer choice. *Journal of Economic Behavior and Organization* 1(1), 39-60.

Thaler, R. H. 2009. Paying a price for the thrill of the hunt. *New York Times* (November 14), available at: http://www.nytimes.com/2009/11/15/business/economy/15view.html.

Thaler, R. H., E. J. Johnson. 1990. Gambling with the house money and trying to break even: The effects of prior outcomes on risky choice. *Management Science* 36(6), 643-660.

Wooldridge, J. M. 2010. *Econometric analysis of cross section and panel data*. The MIT Press, Cambridge, MA.